# Application of AI in Nutrition

Ritu Ramakrishnan, Tianxiang Xing, Tianfeng Chen, Ming-Hao Lee, Jinzhu Gao

*Abstract*— In healthcare, artificial intelligence (AI) has been changing the way doctors and health experts take care of people. This paper will cover how AI is making major changes in the health care system, especially with nutrition. Various machine learning and deep learning algorithms have been developed to extract valuable information from healthcare data which help doctors, nutritionists, and health experts to make better decisions and make our lifestyle healthy. This paper provides an overview of the current state of AI applications in healthcare with a focus on the utilization of AI-driven recommender systems in nutrition. It will discuss the positive outcomes and challenges that arise when AI is used in this field. This paper addresses the challenges to develop AI recommender systems in healthcare, providing a well-rounded perspective on the complexities. Real-world examples and research findings are presented to underscore the tangible and significant impact AI recommender systems have in the field of healthcare, particularly in nutrition. The ongoing efforts of applying AI in nutrition lay the groundwork for a future where personalized recommendations play a pivotal role in guiding individuals toward healthier lifestyles.

*Index Terms*— Artificial Intelligence, Machine Learning, Vectorization, Recipe Recommendation Systems, Nutrition recommendations, Nutrition, Health care, recommendation system, Natural language Processing.

## I. INTRODUCTION

Applications of Artificial Intelligence (AI) in nutrition are growing rapidly and will grow even bigger in the upcoming decade. AI needs a large amount of data to make decisions and predictions. It evolves with learning, reasoning, adapting, and problem-solving. AI has been used in specialized systems to develop evidence-based nutrition information from larger databases and provide best practices. It can give someone quick answers to complex nutrition problems and personalize nutrition according to lifestyles. The integration of AI technologies, such as machine learning, deep learning, and natural language processing, has emerged as a powerful force in helping satisfy the nutritional needs of the people. It also helps to detect early signs of disease, optimize treatment plans, and improve the overall patient experience. AI provides a detailed examination of their body using the applications.

AI applications have already been developed to help the public to make healthier food choices. For example, various mobile applications have been created using AI algorithms to study eating habits and provide meal plan recommendations based on their preference for their profile. These applications also track the user's progress to ensure that the advice is effective for their lifestyle.

In current society, everyone has poor dietary habits, which lack nutritional benefits. According to the recent results of the WHO (World Health Organization), approximately 41 million people die due to non-communicable diseases in their body [1]. Nutrition could play a role in helping those with such ailments. To do so with the current technology available we are proposing a recommendation system for their day-to-day lifestyle.

The goal of this research paper is to study AI applications in nutrition and address the challenges in making nutrition recommendations to common people. Looking at the future directions will give a better understanding of how AI is changing healthcare. This paper aims to set more groundwork for exploring technology with medicine.

.

## II. RELATED WORK

Prompt Engineering Recent developments in nutrition recommendation systems have seen various innovative approaches, reflecting the growing interest and advancements in this area. Here are some notable examples:

### A. Nutrional Plans Recommender: A Decision Tree Model

Based on the user input information the model proposes a nutritional plan to the users [2].

- Key Features: This model uses the decision tree technique to recommend the nutritional approach. The BMI and BMR are calculated with the information provided by the user. Follow ups with the patient check whether the recommendations have improvements in the patient's lifestyle.

- Decision Tree Modelling: The decision tree model uses the scikit-learn tool to evaluate the accuracy of the tree.

- Dataset: The project gets the data from the users and stores information in the database. The BMI, BMR and macro nutrient for the users are calculated.

### B. Nutrition Deficiency Using Random Forest Model [3]

- Key Features: The model uses the forest tree to subset the classification of disease and predict food diseases.

- Random Forest Model: The model is trained using the random forest, decision tree and KNN. When curving,

Ritu Ramakrishnan, Tianxiang Xing, Tianfeng Chen, and MingHao Lee are equal contribution to the paper. Jinzhu Gao is corresponding author (jgao@pacific.edu). All the authors work with department School of Engineering and Computer Science, University of The Pacific, California, USA.

the validation random forest model has higher accuracy and recall rate.
- Dataset: The food dataset stores the food type, minerals, grams and food disease in different columns. The model predicts the disease with three other columns.

### C. Health status improvement using dietary standards

For patients who suffer from obesity, type II diabetes and hypertension are measured and compared with the changes in their bodies. The patients were divided into three groups with no guidance, guided by a dietitian and guided by a dietitian and a physician. It was found that diet has a significant impact on patient weight and blood pressure observed in the graphs of the guided groups [4].

### D. Seasonal variation and Food recommendation

Seasonal change can impact the food recommendations, the price and availability of the foods. For example, tomatoes, eggplant and cucumber are seasonal and are consumed more in the summer and autumn. Therefore, the seasonal change would affect the intake of nutrition in people's diet [5].

The food recommendation system supports a healthy diet based on the user input and their eating habits. There are various websites that get user input and classify them as healthy or non-healthy. They then suggest the healthiest diet according to their lifestyle and to follow up with the user by asking them to upload the picture of the meal [6].

### E. BMI standard calculations

The nutritional research was conducted in China to check the health and dietary requirements of the people. Having a BMI > 30 implies a weight problem according to the WHO but it is different depending on the countries, and it has larger variations depending upon the specific places [7].

## III. METHODOLOGY

We developed a nutritional AI application that helps our daily day-to-day lifestyle. It is like a personal assistant that helps you achieve your health goals by suggesting nutritional meal plans, exercises, and supplements. The main goal is to create a smart nutrition application that uses AI to provide personalized recommendations. To achieve the goal, we use a structured approach of data collection, model development, and validation. The workflow of the recommendations system is shown in Figure 1.

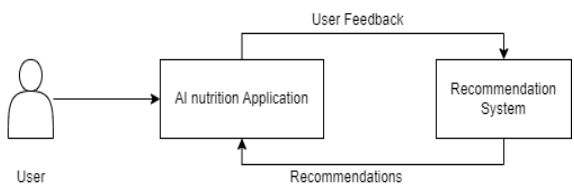

Fig. 1. AI nutrition Application

The system starts with collecting the user information like their height, weight, and health history. A trained model will then use the information to recommend healthy meals, exercise, and nutritional supplements. The following subsections will deep-dive into the specific features of this nutritional application that could benefit the users.

### A. Nutritional Diet Chatbot

The methodological approach of integrating an NLP-driven model into a nutritional chatbot focuses on user interactions, intent extraction, and dynamic data retrieval for personalized dietary recommendations. The main components of the nutrition chatbot application are shown in Figure 2.

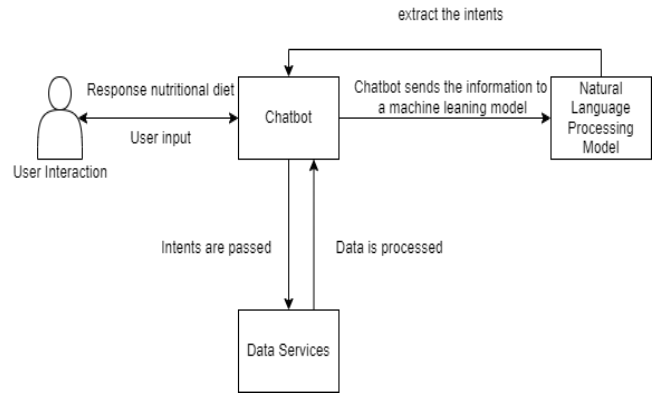

Fig. 2. Nutritional Diet Chatbot

*1) User Interaction and Intent Extraction*

The user provides information about their dietary preferences, restrictions, and health goals through natural language interaction. The information is collected to provide personalized recommendations. These inputs and dietary quires are passed into the dietary model. The intent of the user is then extracted based on the information available.

*2) Dynamic data retrieval*

After the intent extraction, the model dynamically retrieves relevant data from the model. The intents are cross-referenced with the available data services. The data services consist of different datasets including nutritional information, recipe databases, and user-specific profile information. This will also ensure whether the relevant data is extracted for the recommendation.

*3) Dietary recommendation response*

The specific dietary suggestions generated by the application are based on the user's input and dietary type. These recommendations aim to align with the user's preferences, offering a range of meal ideas that cater to their specified diet. The user input is processed for improving the model.

### B. Exercise Recommender

The exercise recommender system is a hybrid system which uses the content-based technique or collaborative technique [8] to recommend the workout routines. The workout routines are crafted based on their user profile and individual fitness goals. The workflow of the nutrition recommender system is shown in Figure 3.

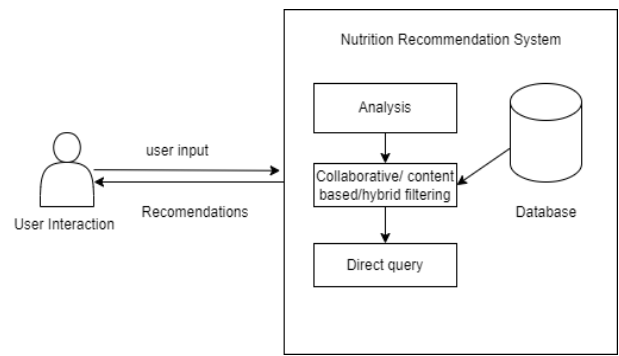

Fig. 3. Nutrition Recommendation System

*1) Understanding Fitness Goals:*

The nutrition application collects the user journey by understanding fitness goals, Body Mass Index (BMI), age, and overall well-being. Through user interactions, individuals share their exercise preferences, past fitness experiences, and any specific health considerations for personalized recommendations.

*2) Machine Learning and Hybrid Intelligence:*

The AI algorithm processes datasets, user inputs, and combines individual preferences, workout histories, and adapts to evolving fitness objectives. After that, it recommends the exercise routine for the user.

*3) Adaptability, Progress Tracking:*

This module is designed to highlight the user's achievements. By continuously assessing performance, it refines routines, ensuring each workout remains challenging for its capabilities. It also suggests exercises with the option of a quick at-home workout or a comprehensive gym session.

C. *Nutritional Supplement Recommender*

The Nutritional Supplement Recommender aims to guide individuals toward the right nutritional supplements available in nearby stores, ensuring the integration of technology and well-being. The nutrition Supplement Recommender model as similar is shown in Figure 3.

*1) User input:*

Users share information about their dietary preferences, health goals, and any specific requirements gathered from their profile, which will be used to suggest the supplements.

*2) Machine Learning Recommendation Model:*

The machine learning model processes enormous amounts of data to learn patterns and make informed suggestions. It considers preferences, health history, and adapts to changes in the wellness journey over time.

*3) Supplement Recommendations:*

The recommender model gives the personalized supplement suggestion based on the user profile. Recommendations of vitamins, minerals, or specific dietary supplements are finely tuned to meet the user's individual health objectives.

*4) Locating Nearby Stores:*

To improve the user experience, the system shows nearby stores that carry the recommended supplements. This ensures accessibility and ease of purchase.

D. *Nutritionist Recommender*

The Nutritionist Recommender takes a personalized turn by integrating content-based recommendations through a machine-learning model. This system aims to guide individuals to nearby nutritionists with top ratings, ensuring expert guidance aligned with the personal health of the users. The nutritionist recommender model as similar is shown in Figure 3.

*1) Understanding User Profile:*

Users share information about their dietary preferences, health challenges, and specific goals they aim to achieve through nutritional guidance.

*2) Machine Learning Model with Content-Based Recommendation:*

The Nutritionist Recommender machine learning model follows the content-based recommendation. This model analyzes user profiles, considering factors such as dietary preferences, health conditions, and past interactions with nutritionists. The model then suggests nutritionists whose expertise aligns closely with the user's needs.

*3) Tailored Nutritionist Recommendations:*

The Recommender suggests a list of nutritionists that aligns with the user profile. These professionals are chosen based on their expertise in addressing specific health concerns, dietary preferences, and positive ratings from other users with similar wellness goals.

*4) Locating Nearby Nutritionists with Best Ratings:*

To ensure convenience and quality, the system identifies nearby nutritionists with the best ratings. By considering user reviews and ratings, the Recommender directs individuals to the nutritionists.

E. *Model Training*

The Nutrition model gives personalized recommendations based on user inputs and dietary preferences. This method ensures that the application considers individual needs, provides accurate dietary suggestions, and contributes to a healthier lifestyle.

*1) Data Collection and Preprocessing:*

Various data on user inputs, including dietary preferences, health goals, restrictions, and historical information for a personalized experience are gathered. The collected data is then cleaned, and standardization processes are used to remove inconsistencies and ensure uniformity in the dataset. The missing values are identified in the dataset.

*2) Feature Engineering:*

Features such as nutritional requirements, dietary restrictions, and taste preferences are carefully gathered to create a personalized user profile.

*3) Algorithm Selection:*

A hybrid approach that uses both machine learning (such as collaborative filtering) and deep learning (neural networks) is chosen to increase the strengths of each, enhancing recommendation accuracy. Reinforcement learning elements are also integrated to allow the model to adapt and refine recommendations based on user feedback and changing preferences.

*4) User-Item Interaction Modeling:*

The model dynamically updates user profiles, capturing real-time interactions and adjustments in dietary habits to provide personalized recommendations that change with the user.

*5) Training Iterations:*

The model undergoes both batch and online training iterations, allowing it to learn from historical data and adapt in real time to changing user inputs and preferences.

The validation metrics including precision, recall, and F1 score, are employed to assess the model's performance, ensuring that recommendations align with dietary guidelines and user expectations.

*6) Fine-Tuning and Hyperparameter Optimization:*

Fine-tuning based on user feedback and iterative optimization of hyperparameters to ensure that the model becomes increasingly adaptable to personalized recommendations.

*F. Results and Personalized Recomendtions*

In this section, we investigate the training results of two distinct models, BERT, and a custom neural network. Each model is finely tuned for specific tasks, highlighting unique learning capabilities. A comparative analysis highlights their strengths and trends.

Fig. 4. BERT Model training Diagram

1) *BERT Model (as shown in Fig. 4):*

    a) *Training Loss and Accuracy Trends:*

    The training loss consistently decreases, highlighting effective learning and improved predictions. Training accuracy starts at 84.80%, maintaining a stable trend and concluding at 83.96% by the 19th epoch.

    b) *Validation Loss and Accuracy Trends:*

    Validation loss mirrors the training loss, indicating effective learning and generalization. Validation accuracy starts at 85.14%, fluctuates significantly, reaching a peak of 84.88%, and ends at 83.93% by the 19th epoch.

2) *Custom Neural Network (As shown in Fig. 5):*

    a) *Training Accuracy and Loss Trends:*

    The model starts with a training accuracy of 81.67%, concluding at 78.66%. This suggests potential overfitting or a need for adjusting the learning rate. Training loss starts at 0.3801, consistently improving to 0.2998, signifying learning and enhanced predictions.

    b) *Validation Accuracy and Loss Trends:*

    Validation accuracy starts higher than training accuracy, indicating the need for a more challenging validation set. It fluctuates more, hinting at struggles in generalization. Validation loss starts at 0.8075 and concludes at 0.7798, showing a consistent improvement.

3) *Comparison:*

    a) *Accuracy and Loss Trends:*

    BERT exhibits stable increases in accuracy and decreases in loss, leveraging its pre-training advantage. The custom neural network's fluctuating performance suggests a need for further fine-tuning and regularization.

    b) *Learning Rate and Epochs:*

    BERT requires fewer epochs for fine-tuning, while the neural network may need more for optimal performance.

    c) *Generalization:*

    BERT excels in generalization due to contextual embeddings. The custom neural network may require more fine-tuning for stability and better generalization.

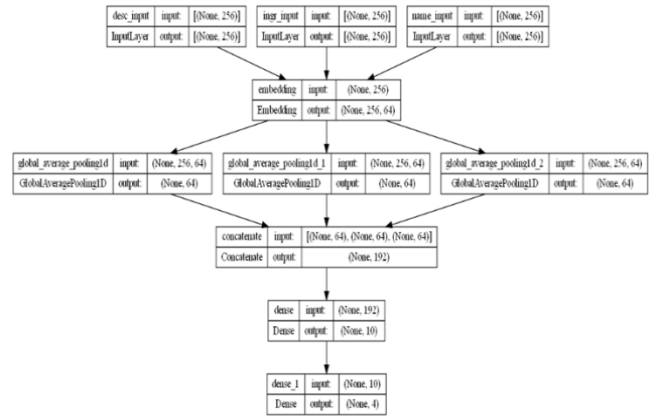

Fig. 5. Neural Network Model Training Diagram

4) *Architecture Model:*

    a) *Inputs:*

    - desc_input: An input layer for descriptions, which accepts sequences of a fixed length (256 in this case).

    - ingr_input: Another input layer, for ingredients, also accepting sequences of length 256.

    - name_input: A third input layer, for the names or titles, with the same sequence length of 256.

    These input layers suggest that the model is designed to handle several types of structured input data, each of a fixed size, which is common in multi-input models where various kinds of data are processed.

    b) *Outputs:*

    - Embedding: This layer outputs a 256x64 matrix, which indicates that each of the 256 sequence elements is mapped to a 64-dimensional embedding.

    - The global_average_pooling1d layers following the embedding layers suggest that the model is reducing the dimensionality of the data, taking the average over the sequence length, and outputting a 64-dimensional vector for each.

    - Concatenate: This layer combines the outputs of the three global average pooling layers, resulting in a 192-dimensional vector (since 3x64=192).

    - The final layers, dense and dense_1, are fully connected layers that output a 10-dimensional vector and a 4-dimensional vector, respectively. These are typically used for classification, with the

dimensions corresponding to the number of classes or labels being predicted.

*5) Personalized Recommendations*

Table I gives an example of how the nutrition app interprets user input, identifies the dietary intent, matches it with an appropriate dietary type, and responds with personalized food recommendations. Each row represents a user scenario, detailing the user input and dietary type and recommends the healthy dietary meal plans according to user profile.

TABLE I. DIETARY MEAL RECOMMENDATIONS

| User Input Intent | Dietary Type | Recommendations |
|---|---|---|
| I'm a vegetarian | Vegetarian | Try our delicious lentil soup or grilled veggie bowl |
| Gluten-free dinner ideas | Gluten-Free | How about quinoa salad or grilled chicken with veggies |
| Protein-rich snacks | High-Protein | Snack on almonds, Greek yogurt, or a protein smoothie |
| Looking for keto meals | Keto | Feast on avocado and bacon salad or salmon with veggies |

Table II explains how the user input is calculated as the fitness goal intents, and how it is converted into the recommendations.

TABLE II. EXERCISE RECOMMENDATIONS

| Finest Goal | User Input | Recommendations |
|---|---|---|
| Weight Management | I want to lose weight and tone my body. | Personalized Hybrid HIIT incorporating bodyweight exercises and nutritional guidance based on weight management goals |
| Muscle Building | I aim to build muscle mass. | Strength Training with a hybrid approach, considering both user preferences and targeted muscle groups |
| Cardiovascular Health | Improving my cardiovascular fitness. | Hybrid Cardio Workouts tailored to the user's preferred activities, considering both intensity and variety |
| Quick Workouts | Short and effective workout ideas. | Dynamic 20-minute Hybrid Circuit Training combining both cardio and strength exercises for time-efficient effectiveness |

Table III outlines personalized nutritional supplement recommendations generated by the Nutritional Supplement Recommender based on the user's health goals. Each row represents a user scenario, detailing the health goal, the recommended supplement, and the nearby store where the supplement can be conveniently purchased.

Table IV provides personalized recommendations for nutritionists with top ratings in nearby locations. Each row represents a user scenario, detailing the health goal, dietary preferences, the recommended nutritionist, their location, and the ratings received. The table serves as a valuable guide for individuals seeking expert nutrition guidance, ensuring that the recommended professionals are both nearby and highly rated by their peers.

TABLE III. NUTRITIONAL SUPPLEMENT RECOMMENDATIONS

| User Goal | Recommended Supplement | Nearby Store |
|---|---|---|
| Immune Support | Vitamin C + | Health Haven Pharmacy |
| Muscle Recovery | BCAA (Branched-Chain Amino Acids) | Fitness Essentials Store |
| Bone Health | Calcium + Vitamin D. | Natural Wellness Market |
| Energy Boost | MCT Oil Capsules | Keto Fuel Superstore |

TABLE IV. NUTRITIONIST RECOMMENDATIONS

| User Goal | Dietary Preferences | Recommended Nutritionist | Ratings | Location |
|---|---|---|---|---|
| Weight Management | Vegetarian | Dr.Sarah Wellness Expert | 4.9 | Nutri Care Center |
| Sports Nutrition | High-Protein | Fitness Fuel Nutrition Studio | 4.8 | Athletic Health Hub |
| Stress Management | Balanced Diet | Eating Therapist | 4.7 | Nutrition Spa |
| Energy Boost | Plant-Based | Vegan Nutrition | 4.7 | Green Life Wellness Center |

Tables I, II, III, and IV encapsulate the personalized recommendations from the nutrition app, spanning dietary preferences, exercise routines, nutritional supplements, and recommended nutritionists. Below is a scenario where these recommendations converge to provide a comprehensive health and well-being guide.

*a) User Scenario:*

Jane, a health-conscious individual, inputs her preferences into the nutrition application.

*b) Dietary Recommendations (Table I):*

Jane states, "I'm a vegetarian.". The application recommends trying the delicious lentil soup or grilled veggie bowl.

*c) Exercise Recommendations (Table II):*

Jane expresses a fitness goal, "I want to lose weight and tone my body." The application suggests a personalized Hybrid HIIT, and nutritional guidance based on her weight management goals.

*d) Nutritional Supplement Recommendations (Table III):*

For immune support, the app recommends Vitamin C+ from Health Haven Pharmacy.

*e) Nutritionist Recommendations:*

Considering her weight management goal and vegetarian preferences, the app recommends Dr. Sarah, a Wellness Expert, with a high rating of 4.9 at Nutri Care Center.

This integrated approach ensures that Jane receives a holistic set of recommendations tailored to her individual needs. The nutrition app seamlessly combines insights from the BERT model, custom neural network, and architecture model, offering Jane a personalized roadmap to health and well-being. This consolidated view provides a detailed analysis of model outcomes and demonstrates the practical application of these findings in offering personalized health and wellness recommendations.

## IV. LIMITATIONS AND FUTURE WORKS

There are still some limitations in the system that we plan to resolve in the future.

### A. Limitations

*1) Data Availability and Quality:*

The effectiveness of our recommendation systems depends on the availability and quality of data. Limited data or inaccuracies in user input may impact the precision of recommendations. The users also need to give the correct information to get accurate results.

*2) Nearby Store Information:*

For the Nutritional Supplement Recommender, the accuracy of nearby store recommendations is challenging on real-time data updates. Changes in store inventory or closures may affect the availability of recommended supplements.

*3) Expertise Scope:*

In the Nutritionist Recommender, the scope of expertise covered by the content-based recommendation model may have limitations. Incorporating a broader range of specialties and continuously updating the expertise database is crucial.

*4) Data Privacy:*

In the current real-world scenario, data privacy is a critical concern when it comes to maintaining databases and safeguarding user information. Organizations handling sensitive data, especially in fields like healthcare, need to ensure the security and privacy of the data they collect.

### B. Future Works

*1) Hybrid Recommendation Models:*

The hybrid recommendation models combine collaborative filtering, content-based filtering, and other emerging techniques to further enhance the precision of recommendations for a better user experience.

*2) Real-Time Data Integration:*

Implementing real-time data integration for the Nutritional Supplement Recommender to ensure up-to-date information on supplement availability in nearby stores.

*3) User Interaction*

Develop a more sophisticated user feedback system to collect detailed insights on the effectiveness and user satisfaction with recommended exercises, diets, or supplements. Exploring innovative strategies to enhance user engagement, motivation, and adherence to recommended fitness and nutrition plans, fostering long-term lifestyle changes.

*4) Expanded Nutritionist Database:*

Expanding the nutritionist database with a wider array of specialties and continuously updating the model to encompass evolving nutritional trends and expertise.

*5) Accuracy:*

User-recommended diets and supplements should not contain allergen items. The recommendations need to be checked with the nutritionist to check whether the recommendations are suitable for the users.

## V. CONCLUSION

In conclusion, the implementation of our AI-driven machine learning model emerges as a promising avenue for nutrition. The focus is to ensure that personalized nutritional guidance reaches every individual, transcending barriers and promoting the well-being of the users. The AI nutrition application serves as an awareness, enlightening individuals about the importance of nutrition and its impact on health.

Using the AI application, the goal is to bridge the informational gap, empowering individuals to make informed choices about their dietary habits, exercise routines, and nutritional supplement needs for their healthy lifestyle. The recommendations extend to making essential nutritional supplements accessible to all users and simplify the process of obtaining these crucial elements for a healthy life. AI-driven machine learning model represents a pivotal step towards a future where nutritional knowledge is accessible to all users.